\documentclass[letterpaper, 10pt, conference]{ieeeconf}
\IEEEoverridecommandlockouts
\usepackage{graphicx}
\usepackage{multirow}
\usepackage{amsmath}
\usepackage{balance}

\title{\LARGE \bf
Towards Trustworthy Audio Deepfake Detection: A Systematic Framework for Diagnosing and Mitigating Gender Bias
}

\author{
    Aishwarya Fursule$^{1}$,
    Shruti Kshirsagar$^{1}$,
    and Anderson R. Avila$^{2,3}$%
    \thanks{$^{1}$School of Computing, Wichita State University, Wichita, KS, USA}%
    \thanks{$^{2}$Institut national de la recherche scientifique (INRS-EMT), Montreal, QC, Canada}%
    \thanks{$^{3}$INRS-UQO Mixed Research Unit on Cybersecurity, Gatineau, Canada}%
    \thanks{Emails: \{axfursule@shockers.wichita.edu, shruti.kshirsagar@wichita.edu,
    anderson.avila@inrs.ca\}}
}

\begin{document}

\maketitle
\thispagestyle{empty}
\pagestyle{empty}

\begin{abstract}
Audio deepfake detection systems are increasingly deployed in high-stakes security applications, yet their fairness across demographic groups remains critically underexamined. Prior work measures gender disparity but does not investigate where it comes from or how to fix it systematically. We present the first diagnosis-first framework that identifies bias source before applying targeted mitigation, evaluated on two models, AASIST~\cite{c7} and Wav2Vec2+ResNet18~\cite{c9}, on ASVSpoof5~\cite{c4}. Our diagnosis shows that bias does not stem from imbalanced training data but from acoustic representation differences, gender leakage in learned features, and structural evaluation asymmetry. We test mitigation strategies across in-processing,  post-processing and combined families, including novel methods introduced in this work. Adjusting the decision threshold separately per gender reduces unfairness by 54–75\% at no cost to detection accuracy, and our new epoch-level fairness regularisation method outperforms existing per-batch approaches. Adversarial debiasing succeeds only when gender leakage is localised, and fails when it is diffuse, an outcome correctly predicted by our diagnosis before training. No single method fully closes the fairness gap, confirming that bias sources must be identified before fixes are applied and that fairer benchmark design is equally important.
\end{abstract}

\section{INTRODUCTION}
\label{sec:intro}

Advances in generative AI have made synthetic speech perceptually
indistinguishable from a genuine human voice, enabling identity
theft, telephony fraud, and audio-based misinformation ~\cite{c1,c2}.
Anti-spoofing countermeasures have progressed from GMM classifiers
with hand-crafted feature representations, through end-to-end models such as
RawNet2~\cite{c6} and AASIST~\cite{c7}, to self-supervised
representations (WavLM~\cite{c8}, Wav2Vec~\cite{c9}) that achieve
state-of-the-art generalisation across unseen attacks~\cite{c10}.

Despite this progress, a critical dimension remains underexamined:
whether these systems perform equitably across demographic
subgroups. Speech signals naturally differ between male and female
speakers in pitch, vocal range, and speaking patterns~\cite{c12},
and models trained without accounting for these differences yield
group-dependent performance~\cite{c13,c29, c30}. Two independent
studies reported higher detection accuracy for
female voices ~\cite{c14,c15} , while~\cite{c16} showed systematically higher false
positive rates for male speakers across six detectors. Our prior
work~\cite{c18} confirmed statistically significant gender bias in
AASIST and four other models on ASVSpoof5 across five fairness
metrics. However, no prior study has identified the specific sources
driving this bias, nor compared pre-processing, in-processing, and
post-processing mitigations on the same model and dataset.

Existing interventions are applied ad hoc, without confirming
which bias sources are active. As we demonstrate, applying sample
reweighting to a gender-balanced dataset not only fails to improve
fairness but actively degrades accuracy. A diagnosis-first approach is necessary; the observed gender disparity does not
arise from training data imbalance but from acoustic
representation differences and structural evaluation protocol
asymmetry. We evaluate two architectures throughout this paper.
Model~1 is AASIST~\cite{c7}, an end-to-end
spectro-temporal graph attention network that processes raw
waveforms directly. Model~2 is
Wav2Vec2-large+ResNet18~\cite{c9}, which pairs a self-supervised
speech front-end with a convolutional back-end classifier.

The main contributions of this paper are:
\begin{enumerate}
    \item We present the first detailed three-level gender bias
    diagnosis for audio deepfake detection, covering data, model,
    and decision sources, evaluated on both Model~1 and Model~2.
    \item We introduce three novel mitigation methods: EAFR,
    SHAP-Guided Feature Suppression (SGFS), and Gender-Neutral
    Embedding Alignment (GNEA).
    \item We provide the first comprehensive comparison of bias reduction strategies,  in-processing, pre-processing,  post-processing, and combined pipelines on both models under identical conditions.
\end{enumerate}

The remainder of this paper is organized as follows:
Section~\ref{sec:related} reviews related work.
Section~\ref{sec:framework} introduces the proposed framework,
detailing both the bias source diagnosis procedure and the
mitigation strategies. Section~\ref{sec:setup} describes the
experimental setup. Section~\ref{sec:results} reports and
discusses results. Section~\ref{sec:conclusion} concludes.

\section{RELATED WORK}
\label{sec:related}

Fairness in audio deepfake detection has only recently begun to
receive dedicated attention. \cite{c14} evaluated machine
learning and deep learning models on a gender-balanced dataset,
reporting higher detection accuracy for female voices and
providing early evidence that gender characteristics influence
detection performance. Further, \cite{c15} observed that models
trained on female audio outperformed those trained on male audio,
linking this to high-pitched artifacts in synthetic speech.
Building on this, \cite{c16} introduced FairSSD, a structured
framework across six detectors, confirming that male speakers
face systematically higher false positive rates across gender,
age, and accent groups. This study \cite{c17} broadened the scope, showing
disparities extend to language and accent dimensions and amplify
with demographically imbalanced corpora. Most directly relevant
to this work, \cite{c18} conducted the first systematic gender
fairness evaluation of AASIST on ASVSpoof5, confirming persistent
disparities across five metrics hidden by aggregate EER. Despite
this diagnostic progress, mitigation remains largely underexplored.
FairVoice \cite{c14} is the only exception, applying
fairness-aware fine-tuning across AASIST, RawNet2, and
Res-TSSDNet, though gender was not its primary target and no
cross-family comparison was conducted. Concurrent with our work,
AFSS \cite{c21} links disparities to domain shift- a finding our
diagnosis independently confirms and extends to the gender dimension. A complete source diagnosis followed by cross-family
mitigation comparison for gender fairness in audio deepfake
detection has not been previously addressed.

Beyond audio, the broader AI fairness literature establishes
three mitigation families: pre-processing, in-processing, and
post-processing~\cite{c22,c23}, with gender bias documented as
amplifying through training when not addressed~\cite{c13}.
However, \cite{c19} proved that standard ERM amplifies
representation disparity over time and proposed distributionally
robust optimization as a remedy, directly motivating our EAFR
design. In \cite{c20}, authors validated that integrated
three-family pipelines outperform single-stage interventions
across six benchmarks. Beyond model architecture, external signal-level factors such as noise and speech enhancement have also been shown to affect deepfake detection performance~\cite{c31}.In visual deepfake detection specifically,
gender-balanced data construction~\cite{c25}, fairness-aware
losses~\cite{c26}, disentanglement learning~\cite{c27}, and
systematic bias auditing~\cite{c32, c33} have all been explored, yet
none of these strategies have been transferred to audio.
Post-processing threshold calibration has not been studied in
either deepfake domain. This gap, alongside the absence of a systematic source diagnosis, motivates the two-stage framework proposed in Section \ref{sec:framework}.

\section{PROPOSED FRAMEWORK}
\label{sec:framework}

\begin{figure}[t]
    \centering
    \includegraphics[width=\columnwidth]{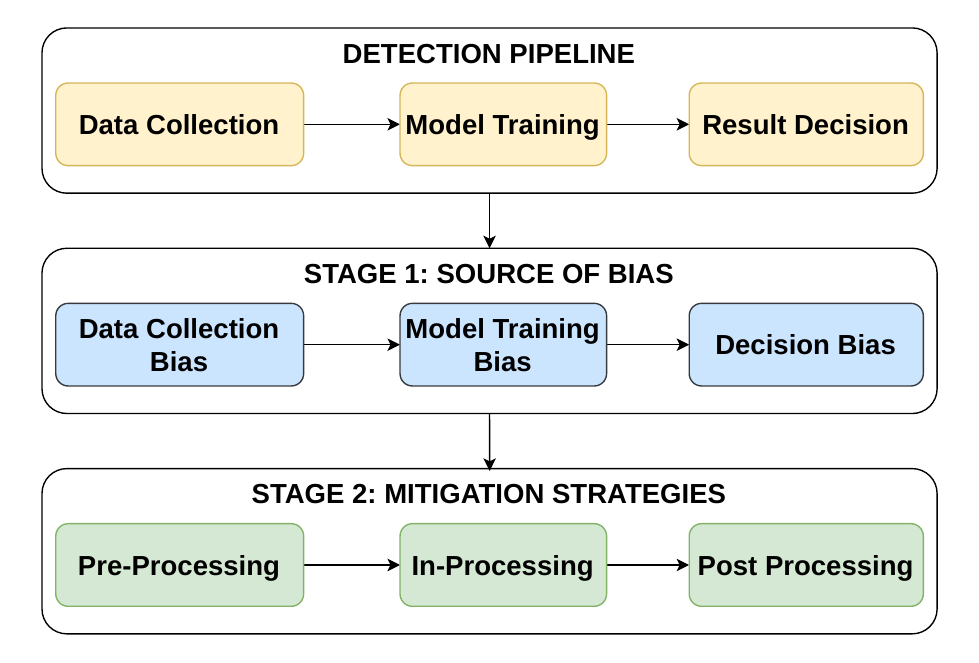}
    \caption{Source-to-mitigation pipeline mapping confirmed bias
    sources at data, model, and decision levels to pre-processing,
    in-processing, and post-processing interventions.}
    \label{fig:pipeline}
\end{figure}
In this section, we describe our proposed two-stage framework for bias diagnosis and mitigation. Fig.~\ref{fig:pipeline} illustrates the proposed two-stage
framework. 
\begin{figure}[t]
    \centering
    \includegraphics[width=\columnwidth]{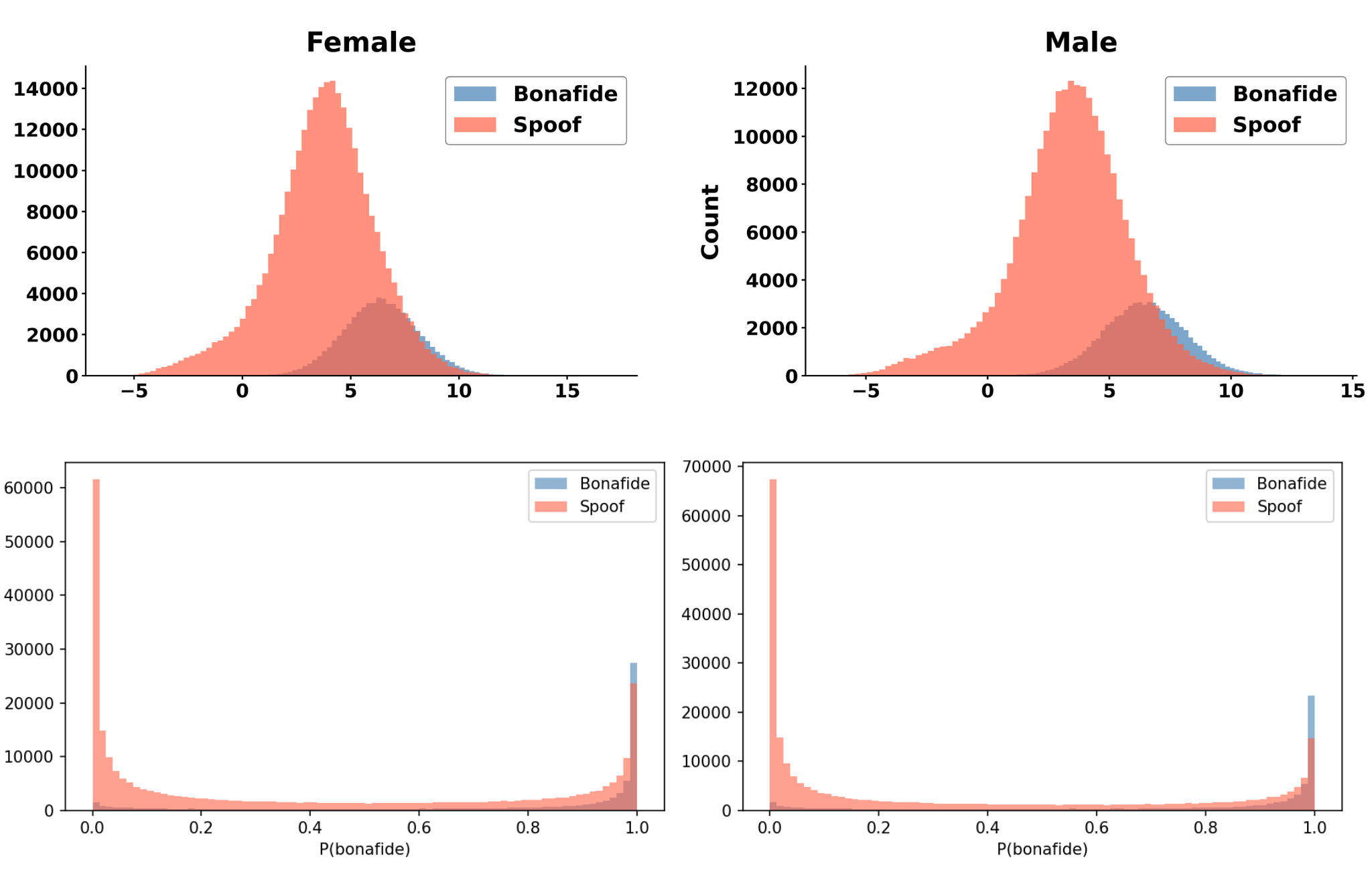}
    \caption{Score distributions per gender on ASVSpoof5.
   Top row (Model~1 / AASIST): 
  Bottom row (Model~2 / Wav2Vec2+ResNet18)}
    \label{fig:scores}
\end{figure}
\subsection{Stage 1: Bias Source Diagnosis}
\label{sec:diagnosis}
\begin{figure*}[t]
    \centering
    \includegraphics[width=\textwidth]{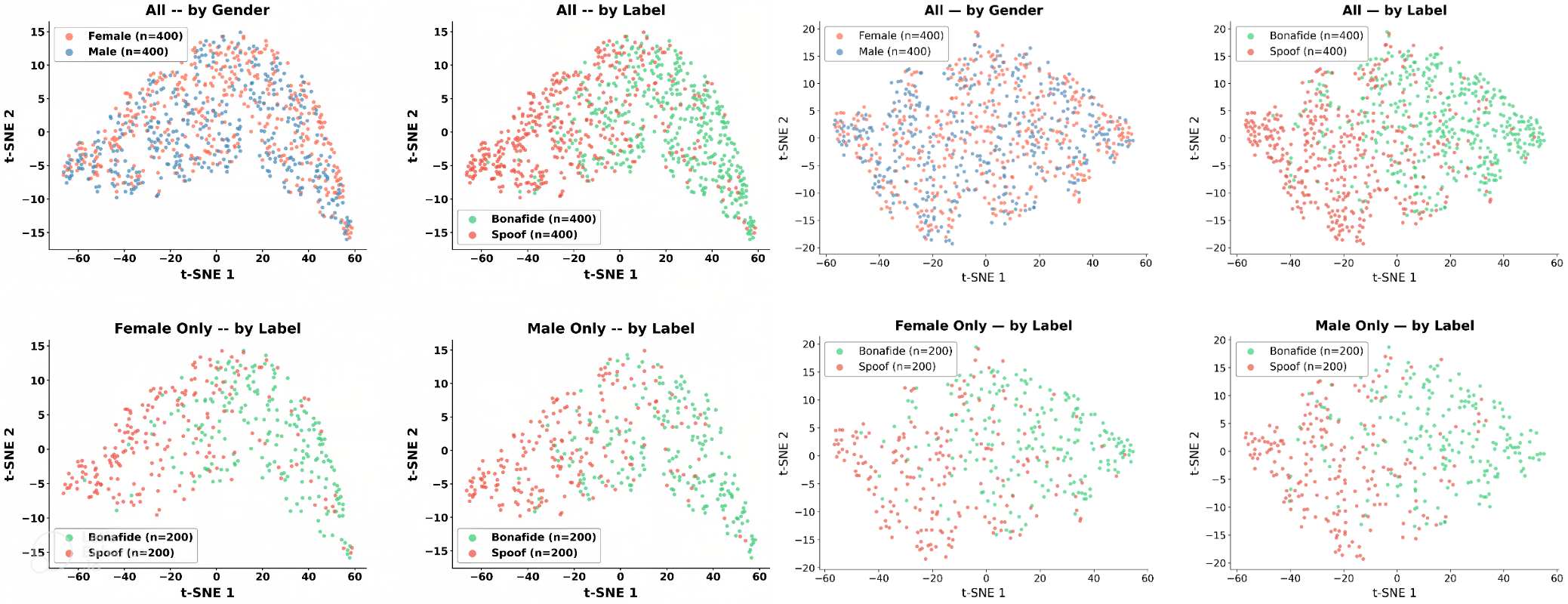}
    \caption{t-SNE projections of 800 embeddings for Model 1
(left two columns) and Model 2 (right two columns), coloured by gender 
and label}
    \label{fig:tsne}
\end{figure*}
In Stage 1, we run eight checks to identify issues at the data, model, and decision levels for both Model 1 and Model 2. 

\subsubsection{Data-Level Checks}
Three checks are performed at the data level. First, we check for gender imbalance in the training data using a chi-squared test to see if reweighting is needed. Second, we examine whether the ratio of real (bonafide) and spoofed samples differs by gender, which could affect error rates. Third, we check the evaluation data using the same test and also look at the distribution of attack types for each gender to identify any imbalance in the dataset.

\subsubsection{Model-Level Checks}
Three checks are used to understand how the model behaves internally. First, we measure score distribution differences to see if the model gives different confidence scores to male and female spoof samples. We compute the difference between the average score for real and spoof samples for each gender. A smaller difference means the model is more confused, and a gap between genders shows bias. Second, we check embedding gender leakage by training a classifier on the model’s features to predict gender. If the accuracy is much higher than 50\%, it means the model’s features contain gender information. Third, we use SHAP to find which parts of the features carry gender information. This helps us see whether the information is concentrated in a few features or spread across many. If it is concentrated, it can be easier to fix; if it is spread out, it is harder to remove. We also use t-SNE plots to visually check how gender and labels are separated in the feature space.

\subsubsection{Decision-Level Checks}
Two checks examine the model’s decision boundary. First, we study training objective bias to see if using binary cross-entropy leads to different false positive rates for males and females at the default threshold. Second, we measure single-threshold bias by comparing the optimal thresholds for each gender. 

\subsection{Stage 2: Mitigation Strategies}
\label{sec:mitigation}

Here, we describe the Stage 2 mitigation strategies.
\subsubsection{Pre-Processing: Sample Reweighting (S1)}
We compute inverse-frequency weights for each gender and label group and use them during training. This method helps address imbalance in the training data. 

\subsubsection{In-Processing: Fairness-Aware Loss (S2)}
This method addresses bias in the training objective by adding a fairness penalty to the standard binary cross-entropy loss:
\begin{equation}
\mathcal{L} = \mathcal{L}_{\mathrm{BCE}} +
\lambda_{\mathrm{fair}}\bigl(
|\mathrm{FPR}_F - \mathrm{FPR}_M| +
|\mathrm{FNR}_F - \mathrm{FNR}_M|\bigr)
\label{eq:fairloss}
\end{equation}
where $\lambda_{\mathrm{fair}} = 0.1$. The false positive rate (FPR) and false negative rate (FNR) for each gender are estimated from the model’s predicted probabilities within each batch.

\subsubsection{In-Processing: Adversarial Debiasing (S3)}
This method reduces gender information in the model embeddings by
adding a Gradient Reversal Layer (GRL)~\cite{ganin2015unsupervised}
between the embedding and a gender classifier. This forces the model
to learn features that do not contain gender information. The scale
parameter is set to $\lambda_{\mathrm{adv}} = 0.05$.

\begin{equation}
\mathcal{L} = \mathcal{L}_{\mathrm{spoof}} +
\mathcal{L}_{\mathrm{gender}}
\label{eq:advloss}
\end{equation}

This method works best when gender information is concentrated in a
few features. If the information is spread across many features, it is
harder for the model to remove.

\subsubsection{In-Processing: Epoch-Accumulated Fairness Regularisation (EAFR)}
EAFR improves S2 by addressing the limitation of small batch sizes.
With a batch size of 24, each gender-label group has only about
6 spoof samples per batch, which is not enough to reliably estimate
fairness metrics such as FPR. To address this, EAFR collects predictions for each group across the
entire epoch before computing the fairness penalty in Eq. 1, using $\lambda_{\mathrm{fair}} = 0.5$, as epoch-level 
estimates are substantially less noisy than per-batch ones, making a 
stronger penalty stable rather than destabilising.

\subsubsection{Post-Processing: Threshold Calibration (TC)}
TC addresses single-threshold bias by using separate thresholds
for each gender. These thresholds are computed using the
equal-error-rate on the development set and applied during
inference. This method does not require retraining. 

\subsubsection{Post-Processing: SHAP-Guided Feature Suppression
(SGFS)}
SGFS reduces gender information in the embeddings by removing
specific features. The dimensions identified by SHAP as
gender-sensitive are set to zero during inference. This method works well when the gender information is
concentrated in a few dimensions. 

\subsubsection{Post-Processing: Gender-Neutral Embedding
Alignment (GNEA)}
GNEA reduces gender information in the embeddings by aligning
values instead of removing them. For the dimensions identified
by SHAP, each value is replaced with the average of the male
and female values:

\begin{equation}
e_d \leftarrow \frac{\bar{e}_{d}^{F} + \bar{e}_{d}^{M}}{2}
\quad \forall\, d \in \mathcal{D}_{\mathrm{SHAP}}
\label{eq:gnea}
\end{equation}

This keeps the overall scale of the embedding while reducing
the gender-specific information in those dimensions.

\subsubsection{Combined Two-Stage Pipelines}
Each in-processing method is also combined with TC. In-processing changes the training process or learned features,
while TC adjusts the decision threshold. These methods target
different types of bias, so using them together can provide additional improvements~\cite{c20}.

\section{EXPERIMENTAL SETUP}
\label{sec:setup}
In this section, we describe the experimental setup, including the dataset, model architecture, and evaluation protocol.
\subsection{ASVSpoof5 Dataset}
All experiments use ASVSpoof5~\cite{c4}, the fifth edition of the ASVSpoof Challenge. Unlike earlier editions, it provides a near-balanced distribution of male and female speakers, making it well-suited for studying gender fairness in audio deepfake detection. The dataset covers both bona fide and AI-generated speech across a wide range of text-to-speech and voice conversion systems. We follow the official evaluation protocol with pre-defined training, development, and evaluation splits, where each split contains entirely different attack types: A01-A08 for training, A09-A16 for development, and A17-A32 for evaluation. This non-overlapping design directly tests generalisation to unseen attack types.

\subsection{Model Architectures}
Two countermeasure systems are evaluated. Model~1
(AASIST~\cite{c7}) processes raw waveforms end-to-end via a
spectro-temporal graph attention network, producing a
160-dimensional embedding passed to a linear binary classifier.
Model~2 (Wav2Vec2-large+ResNet18~\cite{c9}) pairs a
self-supervised Wav2Vec2-large front-end feature extractor with a
ResNet18 back-end, where frame-level representations of shape
$(T, 1024)$ are treated as single-channel two-dimensional feature
maps and classified via an adaptive average pooling head. 

\subsection{Evaluation Protocol}
 We evaluate using  following fairness metrics: 
 The task is formulated as a binary classification problem, with $Y=1$ denoting AI-generated speech, and $Y=0$ representing bonafide speech. The predicted class is denoted by $\hat{Y}$, where $g \in \{f, m\}$ denotes female ($f$) and male ($m$) speakers.  These are derived from classification outcomes such as True Positives (TP), True Negatives  (TN), False Positives (FP), and False Negatives (FN) per gender group.

\begin{enumerate}
\item {Equal Error Rate (EER)} is the primary 
detection performance metric, defined as the point where 
the false acceptance rate equals the false rejection rate. 
A lower EER indicates better overall detection performance:
\begin{equation} 
\mathrm{EER} : \mathrm{FAR}(\theta) = \mathrm{FRR}(\theta)
\end{equation}
where $\theta$ is the decision threshold derived from the 
development set, and the EER gap between genders is:
\begin{equation}
\mathrm{EER\ gap} = \mathrm{EER}_f - \mathrm{EER}_m
\end{equation}

\item {False Positive Rate (FPR) Difference ($d_{\mathrm{FPR}}$)} 
measures the gap in false positive rates between female and male speakers, where a positive value indicates that 
Female bona fide speech is more likely to be incorrectly 
flagged as spoofed:
\begin{equation}
d_{\mathrm{FPR}} = \frac{FP_f}{FP_f + TN_f} - 
\frac{FP_m}{FP_m + TN_m}
\end{equation}

    \item Statistical Parity Difference (SPD) checks whether the model detects spoofed speech at the same rate for both genders, regardless of whether the predictions are correct, 

    \begin{equation}
\mathrm{SPD} = \left| P(\hat{Y}=1 \mid G=f) - 
P(\hat{Y}=1 \mid G=m) \right|
\end{equation}

    \item Equal Opportunity (EOP) goes a step further by focusing only on actual spoof samples, measuring whether the model is equally good at catching spoofed speech from male and female speakers. Equalized Odds extends this further by requiring the model to perform equally well on both spoof and bona fide samples across genders, ensuring fairness in both directions of prediction. 
\begin{equation}
\mathrm{EOP}_g = P(\hat{Y}=1 \mid Y=0, G=g)
= \frac{FP_g}{FP_g + TN_g}
\end{equation}

    \item Predictive Parity Difference (PPD) examines whether a positive prediction flagging a sample as a deepfake is equally reliable regardless of the speaker's gender
\begin{equation}
\mathrm{PPV}_g = P(Y=1 \mid \hat{Y}=1, G=g)
\end{equation}
    
    \item Treatment Equality Difference (TED) looks at the balance between false acceptances and false rejections for each gender group.
    \begin{equation}
\mathrm{TE}_g = \frac{FP_g}{FN_g}
    \end{equation}

\end{enumerate}

\section{RESULTS AND DISCUSSION}
\label{sec:results}
\begin{table*}[t]
\caption{Results on Stage 1: Three-level bias source diagnosis for Model~1 (AASIST)
and Model~2 (Wav2Vec2+ResNet18) on ASVSpoof5.
}
\label{tab:diagnosis}
\centering
\small
\setlength{\tabcolsep}{4pt}
\begin{tabular}{@{}llllll@{}}
\hline
\textbf{Level}
  & \textbf{Source}
  & \textbf{Model~1}
  & \textbf{Status}
  & \textbf{Model~2 }
  & \textbf{Status} \\
\hline
\multirow{3}{*}{Data}
  & Training imbalance
  & $\chi^2\!=\!1.196$, $p\!=\!0.274$
  & Ruled out
  & $\chi^2\!=\!1.196$, $p\!=\!0.274$
  & Ruled out \\
  & Eval protocol asymmetry
  & $\chi^2\!=\!275.13$, $p\!<\!10^{-61}$
  & Confirmed
  & $\chi^2\!=\!275.13$, $p\!<\!10^{-61}$
  & Confirmed \\
  & Attack non-overlap
  & A01-A08 vs.\ A17-A32
  & Confirmed
  & A01-A08 vs.\ A17-A32
  & Confirmed \\
\hline
\multirow{3}{*}{Model}
  & Score separation gap
  & F\,=\,2.713, M\,=\,3.120 (gap\,=\,0.407)
  & Confirmed
  & F\,=\,0.388, M\,=\,0.422 (gap\,=\,0.034)
  & Confirmed \\
  & Gender leakage accuracy
  & 62.5\% ($+$12.5\,pp above chance)
  & Confirmed
  & 53.4\% ($+$3.4\,pp above chance)
  & Weak \\
  & SHAP leakage type
  & Dims 125, 36, 90  - Localised
  & Confirmed
  & Dims 426, 510, 141  - Diffuse
  & Weak \\
\hline
\multirow{2}{*}{Decision}
  & Single threshold bias
  & F\,=\,5.249, M\,=\,5.091, gap\,=\,0.158
  & Confirmed
  & F\,=\,0.773, M\,=\,0.671, gap\,=\,0.102
  & Confirmed \\
  & Training objective bias
  & $d_{\mathrm{FPR}}\!=\!+0.050$
  & Confirmed
  & $d_{\mathrm{FPR}}\!=\!+0.057$
  & Confirmed \\
\hline

\end{tabular}
\end{table*}

\begin{table*}[t]
\centering
\caption{Results on Stage 2: Gender fairness mitigation for Model~1
(AASIST) and Model~2 (Wav2Vec2+ResNet18) on ASVSpoof5.}
\label{tab:results}
\begin{tabular*}{\textwidth}{@{\extracolsep{\fill}}llcccccccc@{}}
\hline
\textbf{System} & \textbf{Family}
  & \textbf{EER\,F\%} & \textbf{EER\,M\%}
  & \textbf{EER Gap}
  & $\boldsymbol{d_\textbf{FPR}}$
  & \textbf{SPD} & \textbf{EOP} & \textbf{PPD} & \textbf{TED} \\
\hline
\multicolumn{10}{@{}l}{\textit{(a) Model~1: AASIST}} \\
\hline
Baseline                  &  -
  & 24.92 & 21.37 & 3.55
  & $+$0.078 & $-$0.016 & $-$0.007 & $-$0.099 & $-$0.042 \\
$+$ S1 (Reweighting)      & Pre
  & 27.78 & 23.94 & 3.84
  & $+$0.127 & $+$0.031 & $+$0.021 & $-$0.101 & $+$0.053 \\
$+$ S2 (Fairness Loss)    & In
  & 30.88 & 26.53 & 4.35
  & $+$0.110 & $+$0.089 & $+$0.043 & $-$0.094 & $+$0.116 \\
$+$ S3 (Adversarial GRL)  & In
  & \textbf{22.48} & \textbf{19.30} & 3.19
  & $+$0.078 & $+$0.064 & $+$0.054 & $-$0.116 & $-$0.132 \\
$+$ EAFR                  & In
  & 25.75 & 23.22 & \textbf{2.52}
  & $+$0.067 & $+$0.053 & $+$0.040 & $-$0.087 & $-$0.106 \\
$+$ TC                    & Post
  & 24.92 & 21.38 & 3.54
  & $+$0.036 & $-$0.013 & $+$0.035 & $-$0.074 & $-$0.000 \\
$+$ SGFS                  & Post
  & 24.88 & 21.34 & 3.54
  & $+$0.050 & $+$0.029 & $-$0.009 & $-$0.099 & $-$0.041 \\
$+$ GNEA                  & Post
  & 24.88 & 21.35 & 3.53
  & $+$0.050 & $+$0.030 & $-$0.009 & $-$0.098 & $-$0.041 \\
$+$ S3 $+$ TC             & Combined
  & \textbf{22.48} & \textbf{19.30} & 3.19
  & $+$0.063 & $+$0.046 & $-$0.026 & $-$0.116 & $-$0.132 \\
$+$ EAFR $+$ TC           & Combined
  & 25.75 & 23.22 & \textbf{2.52}
  & $+$0.056 & $+$0.041 & $-$0.023 & $-$0.087 & $-$0.106 \\
$+$ SGFS $+$ TC           & Combined
  & 24.88 & 21.34 & 3.54
  & $+$\textbf{0.035} & \textbf{$+$0.012} & $+$0.035 & $-$0.099 & $-$0.041 \\
$+$ GNEA $+$ TC           & Combined
  & 24.88 & 21.35 & 3.53
  & $+$\textbf{0.035} & \textbf{$+$0.012} & $+$0.035 & $-$0.098 & $-$0.041 \\
\hline
\multicolumn{10}{@{}l}{\textit{(b) Model~2: Wav2Vec2-Large + ResNet18}} \\
\hline
Baseline                  &  -
  & 28.43 & 26.95 & 1.47
  & $+$0.054 & $+$0.043 & $-$0.034 & $-$0.067 & $+$1.269 \\
$+$ S1 (Reweighting)      & Pre
  & 31.10 & 29.47 & 1.63
  & $+$0.098 & $+$0.061 & $-$0.038 & $-$0.071 & $+$1.403 \\
$+$ S2 (Fairness Loss)    & In
  & 25.92 & 25.39 & \textbf{0.53}
  & $+$0.057 & $+$0.031 & $-$0.042 & $-$0.056 & $+$0.987 \\
$+$ S3 (Adversarial GRL)  & In
  & 34.02 & 33.65 & 0.37
  & $+$0.057 & $+$0.017 & $-$0.022 & $-$0.030 & $+$0.826 \\
$+$ EAFR                  & In
  & 31.67 & 30.26 & 1.41
  & $+$0.058 & $+$0.058 & $-$0.056 & $-$0.060 & $+$1.550 \\
$+$ TC                    & Post
  & \textbf{28.41} & \textbf{26.95} & 1.46
  & $+$0.014 & $+$0.001 & $-$0.015 & $-$0.041 & $+$0.385 \\
$+$ SGFS                  & Post
  & 28.44 & 26.97 & 1.47
  & $+$0.054 & $+$0.042 & $-$0.034 & $-$0.067 & $+$1.262 \\
$+$ GNEA                  & Post
  & 28.42 & 26.98 & \textbf{1.46}
  & $+$0.054 & $+$0.042 & $-$0.034 & $-$0.067 & $+$1.265 \\
$+$ S2 $+$ TC             & Combined
  & 25.91 & 25.39 & \textbf{0.52}
  & $+$\textbf{0.005} & $-$0.005 & $-$0.005 & $-$0.031 & $+$0.387 \\
$+$ S3 $+$ TC             & Combined
  & 34.02 & 33.65 & 0.37
  & $+$0.004 & $-$0.003 & $-$0.004 & $-$0.026 & $+$0.389 \\
$+$ EAFR $+$ TC           & Combined
  & 31.67 & 30.26 & 1.41
  & $+$0.014 & $+$0.002 & $-$0.014 & $-$0.038 & $+$0.387 \\
$+$ SGFS $+$ TC           & Combined
  & 28.44 & 26.97 & 1.47
  & $+$0.015 & $+$0.001 & $-$0.015 & $-$0.042 & $+$0.390 \\
$+$ GNEA $+$ TC           & Combined
  & 28.42 & 26.98 & \textbf{1.46}
  & $+$0.014 & $+$0.001 & $-$0.014 & $-$0.041 & $+$0.390 \\
\hline
\end{tabular*}
\end{table*}

We report the results for both stages of the proposed framework and discuss the findings in light of existing literature, as described below.

\subsection{Stage 1: Bias Source Diagnosis}
\label{sec:results_diagnosis}

Table~\ref{tab:diagnosis} summarises the full eight-check
diagnosis. 

\subsubsection{Data Level}
The training set is gender-balanced ($\chi^2 = 1.196$,
$p = 0.274$) for both models, which rules out sample
reweighting as a necessary solution. This is important because
prior work in AI fairness~\cite{c22,c23} and voice biometrics~\cite{c14}
often assumes that performance differences are caused by training
imbalance and applies reweighting or oversampling. Our analysis
shows that this assumption does not hold for ASVSpoof5. Therefore,
using reweighting would not be appropriate, which we later confirm
experimentally in Stage~2. Although the training data is balanced, the evaluation set shows
a strong imbalance ($\chi^2 = 275.13$, $p < 10^{-61}$). Female
speakers have a higher bonafide-to-spoof ratio ($1{:}4.10$) compared
to male speakers ($1{:}3.71$), and 13.4\% more spoof samples come
from unseen attack types (A17-A32). This creates a bias in the
evaluation setup that affects both models and cannot be fixed by
changing the model. This finding supports earlier work such as AFSS~\cite{c21}, which
shows that fairness issues in audio deepfake detection are linked
to differences between training and evaluation data.  Yadav et al.~\cite{c16} also showed
that evaluation design can cause differences in false positive
rates across groups, and our chi-squared results provide a clear
explanation for this effect.

\subsubsection{Model Level:}

Fig.~\ref{fig:scores} shows the per-gender score distributions for both models. 
For Model~1 (top panel), the spoof score distribution for female speakers is 
centred lower (mean $= 2.713$) than for male speakers (mean $= 3.120$), 
producing a mean score difference of $0.407$, indicating greater difficulty 
distinguishing genuine from synthetic female speech. For Model~2 (bottom panel), 
the same directional asymmetry holds but the gap ($0.034$) is substantially 
smaller, falling below the $0.1$ threshold 
associated with practically significant score asymmetry, consistent with the advantage ~\cite{c8,c9}, yet confirming that pre-training alone does not eliminate score asymmetry. 
This aligns with Bird and Lotfi~\cite{c15}, who linked gender-differential 
accuracy to high-pitched spectral artefacts in female synthetic speech.
Fig.~\ref{fig:tsne} presents t-SNE projections of 800 embeddings coloured 
by gender (top row) and label (bottom row). In both models, female and male 
points are extensively interleaved with no separable clusters~\cite{c13}, 
yet the bottom row reveals greater bona fide--spoof overlap for female 
speakers, directly visualising the elevated female false positive rate 
observed across both models.

\subsubsection{Decision Level}
Both models show a higher false positive rate for one gender
(Model~1: $+0.050$, Model~2: $+0.057$), even though the training
data is balanced. This shows that binary cross-entropy can
introduce bias on its own, which agrees with
Hashimoto~et~al.~\cite{c19}. The optimal thresholds for each gender are also different.
Model~1 has a gap of 0.158 (F: 5.249, M: 5.091) and Model~2 has
a gap of 0.102 (F: 0.773, M: 0.671). This means using a single
threshold puts female speakers at a disadvantage in both models. Per-gender threshold calibration has not been studied before in
audio~\cite{c16,c18} or visual~\cite{c26} deepfake detection,
making this an important finding. Overall, these results show that gender bias comes from multiple
different sources, not just one~\cite{c18,c16}.

\subsection{Stage 2: Mitigation Results}
\label{sec:results_mitigation}
Table~\ref{tab:results} shows the results of all mitigation
methods for both models. The baseline results highlight that
using only one metric, such as EER gap, is not enough to measure
fairness. Although Model~2 has a smaller EER gap than Model~1,
it performs worse on other fairness measures, confirming that
multiple metrics are needed~\cite{c18}. S1 (sample reweighting)
acts as a negative test and makes performance worse for both
models because the training data is already balanced. This
supports the idea that mitigation should only be applied after
confirming the source of bias~\cite{c22}. S2 (fairness-aware
loss) also performs poorly due to unstable estimates from small
batch sizes, although combining it with TC improves results for
Model~2.

S3 (adversarial GRL) shows different behavior for the two
models. For Model 1, where gender information is concentrated
in a few dimensions, S3 is consistent with prior work~\cite{ganin2015unsupervised}. For Model~2, where
leakage is spread across many dimensions, S3 raises EER by approximately six percentage points with negligible fairness benefit. TC is the most
reliable method, reducing false positive rate differences by 54-75\% without affecting accuracy, Combining TC with other
methods further improves fairness, confirming TC as a simple yet effective post-processing intervention \cite{c16,c26}.

SGFS and GNEA reduce bias in Model~1 by targeting specific
dimensions identified by SHAP, and achieve the best results
when combined with TC. For Model~2, these methods have little
effect because the bias is spread across many dimensions.
 Remaining bias is
mainly due to dataset design and evaluation imbalance, which
suggests that future work should also improve benchmark
protocols~\cite{c21}.
\section{CONCLUSION}
\label{sec:conclusion}

This paper addressed gender bias in audio deepfake detection,
a critical yet underexamined threat to the fairness of
anti-spoofing systems. We proposed a diagnosis-first two-stage
framework that systematically identifies active bias sources
before applying mitigation. We evaluated across eight strategies and
three novel methods EAFR, SGFS, and GNEA, on two
architectures. Results show that bias arises from multiple
independent sources: evaluation protocol asymmetry, score
distribution shift, embedding gender leakage, and threshold
bias not training data imbalance. Threshold calibration
reduced $d_{\mathrm{FPR}}$ by 54-75\% at zero accuracy cost,
EAFR outperformed per-batch fairness loss, and adversarial
debiasing succeeded only for Model~1 where leakage was
localised a failure correctly predicted by SHAP before
training. These findings demonstrate two things: that diagnosis must precede mitigation, as applying strategies without identifying confirmed sources of bias actively harms performance; and that future benchmarks must address structural protocol asymmetries to close the fairness gap. This study is limited to a single dataset, ASVSpoof5, with binary 
gender labels; extending to multiple datasets and non-binary gender representations remains important future work.


\addtolength{\textheight}{-3cm}


\end{document}